\documentclass[aps,prl,twocolumn,showpacs,groupedaddress]{revtex4}
\usepackage{graphicx}  % needed for figures
\usepackage{dcolumn}   % needed for some tables
\usepackage{bm}        % for math
\usepackage[english]{babel}
\usepackage{amsfonts,amsmath,amssymb,mathrsfs}
\usepackage{epstopdf}
\usepackage{subfigure}

\usepackage{latexsym}
\def\be{\begin{equation}}
\def\ee{\end{equation}}
\def\bea{\begin{eqnarray}}
\def\eea{\end{eqnarray}}
\def\ben{\begin{enumerate}}
\def\een{\end{enumerate}}

\begin{document}

\title{Covariant Effective Action for Loop Quantum Cosmology from Order Reduction}
\author{Thomas P. Sotiriou}
\affiliation{Center for Fundamental Physics,  University of Maryland, College Park, MD 20742-4111, USA}
\date{\today} 
\begin{abstract}
Loop quantum cosmology (LQC) seems to be predicting modified  effective Friedmann equations without extra degrees of freedom. A puzzle arises if one decides to seek for a covariant effective action which would lead to the given Friedmann equation: The Einstein--Hilbert action is the only action that leads to second order field equations and, hence, there exists no covariant action which, under metric variation, leads to modified Friedmann equation without extra degrees of freedom. It is shown that, at least for isotropic models in LQC, this issue is naturally resolved and a covariant effective action can be found if one considers higher order theories of gravity but faithfully follows effective field theory techniques. However, our analysis also raises doubts on whether a covariant description without background structures can be found for anisotropic models.
\end{abstract}  
\maketitle

According to  the Big Bang scenario the universe appears to emerge from a singularity. However, this conclusion is based on purely classical considerations and  
quantum gravity is expected to provide a more complete understanding of the Big Bang and maybe even do away with the initial singularity. In the framework of a spacetime picture the only way to avoid an isotropic singularity is to have a bounce, in the sense that at a non zero volume the collapse is halted and turned around.

One of the proposals in this direction comes from LQC \cite{Bojowald:2006da}. It has been shown that the cosmological singularity in isotropic minisuperspaces is naturally removed by quantum geometry \cite{Bojowald:2001xe}. The free, massless scalar model, consisting of a flat, isotropic and homogeneous Friedmann-Lema\^itre-Robertson-Walker (FLRW) spacetime sourced by a massless scalar field has been quantized in LQC and can be considered well understood \cite{Singh:2006sg,Ashtekar:2006rx}. Through an effective hamiltonian description it leads to an effective Friedmann equation \cite{Singh:2006sg}\footnote{Remarkably, in the Randall--Sundrum braneworld scenario one gets the same modified Friedmann equation with an opposite sign in front of the $\rho^2$ term \cite{Singh:2006sg}.}
\be
\label{lqf}
H^2=\frac{1}{3}\kappa \rho\left(1-\frac{\rho}{\rho_c}\right),
\ee
where $a(t)$ is the scale factor, $H=\dot{a}/a$,     $\rho$ is the energy density, $\kappa=8\pi\,G$, $G$ is Newtons constant and a dot denotes differentiation with respect to coordinate time. For the full description of the dynamics this modified Friedmann equation should be supplemented with a Klein-Gordon equation for the scalar (or the conservation law $\dot{\rho}=-3H (\rho+p)$ where $p$ is the pressure). This equation remains unmodified as the loop quantization does not affect the matter hamiltonian. Clearly, a bounce occurs when $\rho=\rho_c$. The critical density for this bounce is equal to $\rho_c=\sqrt{3}/(2\pi\kappa\gamma^3 l_p^2)$, where $\gamma$ is the Barbero-Immirzi parameter and $l_p$ is the Planck length. It has been shown numerically and proven rigorously that the effective Friedmann eq.~(\ref{lqf}) reliably describes the evolution of the expectation value of a wave packet in the free massless scalar model \cite{Ashtekar:2006rx,Ashtekar:2006uz,bojoun}. Therefore, despite its limitations, this model and its effective description should be able to give a useful insight into LQC.

A reasonable question to ask is whether the effective modified Friedmann equation  can be derived 
from a diffeomorphism invariant effective action. 
This leads to the following puzzle: Eq.~(\ref{lqf}) is just the usual Friedmann equation with a modified source, {\em i.e.}~there are no extra degrees of freedom. However, we already know that the only covariant action which leads to second order equations under metric variation is the Einstein--Hilbert action and, assuming an FLRW spacetime, these equations yield the usual Friedmann equations. Any modification of the Einstein-Hilbert action would lead to higher order equations under metric variation.

Irrespectively of the uncertainty on this issue at the level of the fundamental underlying theory, covariance is  a standard prerequisite for an effective low energy theory, so the pertinent question is the following: Does the above puzzle imply that a covariant formulation of the effective Hamiltonian leading to eq.~(\ref{lqf}) is not possible, with any implications this might have for the approximations used or the underlining physics, or can we actually simply circumvent that previous argument? This is the question that we want to address here.

One possible answer was  given in Ref.~\cite{Olmo:2008nf}: it was argued that a covariant action can indeed be constructed within the framework of Palatini $f(R)$ gravity (see Ref.~\cite{Sotiriou:2008rp} for a review on $f(R)$ gravity and references therein), which assumes that the connection and the metric are independent variables, but the former does not couple to the matter (unlike metric-affine $f(R)$ gravity \cite{Sotiriou:2006qnplus}). Note that since the only matter considered is a scalar field, couplings between the connection and the matter in the effective theory cannot be probed. This implies that the latter might as well be a metric-affine $f(R)$ gravity, something that passed unnoticed in Ref.~\cite{Olmo:2008nf} \footnote{Palatini $f(R)$ gravity is really a metric theory: the independent connection is essentially just an auxiliary field for which one can solve algebraically and eliminate \cite{Sotiriou:2006qnplus}. The underlying geometry (at least as seen by the matter) is pseudo-Riemannian and this connection can hardly be considered a fundamental field. Therefore, if one wants to consider a theory with non-metricity based on the claim of Ref.~\cite{Olmo:2008nf} that no motivation for assuming a compatibility relation comes from loop quantum gravity, metric-affine gravity is much more suitable.}.

We would like to focus more on another possible way out, which does not even go against conventional wisdom about corrections to the action and variational principle choices. More precisely, we would like to stick to standard metric variation and an action with higher order curvature invariants and propose a different way to do away with the extra degrees of freedom: order reduction. Even though lately higher order theories of gravity, such as $f(R)$ gravity,  have been mostly considered as exact theories, meaning that their field equations are considered as genuinely higher order, it has been long known that another alternative is to treat them as effective field theories, {\em i.e.}~consider the solutions which are perturbatively close to GR as physical ones and the rest as spurious \cite{old} (this technique has also been used recently in Ref.~\cite{DeDeo:2007yn} to ``cure" the viability issues of infrared modification in metric $f(R)$ gravity). 

Which one of the two  is the right way to proceed is not evident by the form of the action, nor is it a matter of choice. It depends on how this action comes about from a more fundamental theory, {\em i.e.} if the extra degrees of freedom can be considered fundamental or are introduced during the derivation of the effective action. A typical, but not the only, example of the second case would be a non-local theory which leads to an effective action with infinite terms and higher order derivatives which produce spurious degrees of freedom. The non-local theory does not have to be fundamental; it might be an effective theory itself derivable from a fundamental theory through a scheme in which locality is lost.

Instead of discarding spurious solutions due to the extra degrees of freedom one by one, the technique of order reduction can be used to derive field equations  which are second order and provide only the physical solutions \cite{old}. So, even  if higher order theories of gravity cannot lead to second order field equations under metric variation if they are considered as exact theories, this is not the case if they are approached as effective theories. In the latter case, they could very well lead to modified Friedmann equations of the sort of eq.~(\ref{lqf}). As we are about to show, one can actually specify the corresponding action (or at least a family of actions) starting from a  Friedmann equation with a modified source, such as eq.~(\ref{lqf}). The need to apply such an effective field theory scheme to find the covariant action may just be a manifestation of the characteristics of the underlying fundamental theory or the way one derives the effective description, such as non-locality.

For simplicity we start from the easiest form of a higher order action, 
 metric $f(R)$ gravity. The action is
\be
\label{action}
S=\frac{1}{2\kappa} \int d^4 x \sqrt{-g} f(R)+S_M(g_{\mu\nu,\psi}), 
\ee
where  $g_{\mu\nu}$ is the metric and $g$ its determinant, $f$ a general function of the Ricci scalar of the metric $R$, $S_M$ denotes the matter action and $\psi$ collectively denotes the matter fields. Variation with respect to $g_{\mu\nu}$ yields
\be
\label{feq}
f' R_{\mu\nu}-\frac{1}{2} f g_{\mu\nu}-[\nabla_\mu \nabla_\nu-g_{\mu\nu}\Box] f'= \kappa\, T_{\mu\nu},
\ee
where $T_{\mu\nu}=-(2/\sqrt{-g})\, \delta S_M/\delta g^{\mu\nu}$ is the stress energy tensor as usual, $\nabla_\mu$ is the metric covariant derivative and $\Box=\nabla^\mu\nabla_\mu$ and the prime denotes differentiation with respect to $R$.
A useful equation is the one obtained by the contraction of eq.~(\ref{feq})
\be
\label{trace}
f' R-2 f+3 \Box f'= \kappa\, T,
\ee
where $T=g^{\mu\nu} T_{\mu\nu}$. Clearly, unlike GR, $R$ is not related algebraically to $T$. Without loss of generality we can parametrize $f$ as 
\be
f(R)=R+2\Lambda+\epsilon \varphi(R).
\ee
The parameter $\epsilon$ is dimensionless and marks the deviation from GR.
One can think of $\varphi$ as a function incorporating all possible corrections to the Einstein--Hilbert action. For instance, if $f$ is thought of as a series expansion then $\varphi=\ldots+a_{-1} l_p^{-4} /R+a_2 l^2_pR^2+ a_3 l_p^{4} R^3+\ldots$, where $l_p$ is the Planck length and the $a_i$ coefficients are dimensionless. 

Since we want to treat metric $f(R)$ gravity, not as an exact but as an effective field theory whose solution ought to be perturbatively close to GR, we have to perform an order reduction to the field equation in order to do away with the spurious degrees of freedom. Note that $\epsilon$ does not exactly play the role of a small parameter here nor does it need to be small. Actually it could easily be absorbed in $\varphi$ and in the series expansion example we gave earlier, it would be redundant as it could be eliminated by a redefinition of the $a_i$'s. However, $\epsilon$ is helpful since it allows us to use the order reduction technique developed in Refs.~\cite{old} (a parameter is needed the vanishing of which leads back to GR). For the whole scheme to be valid of course, $\epsilon\varphi \ll R$ at the range of curvatures considered (here essentially $R\ll l_p^{-2}$, but this can vary according to the application). This can be assumed now and verified a posteriori. At the same time working at order $epsilon$ would essentially mean that we are working to first order in a parameter in which $\phi$ is linear as a correction in an effective action ({\em e.g.}~one of the $a_i$'s).

Due to the form of the field equations, the order reduction amounts to just replacing $R$ and $R_{\mu\nu}$ in order $\epsilon$ terms with the expression one gets for them from the $\epsilon=0$ version of the same equations.
Eq.~(\ref{feq}) reduces to
\begin{align}
\label{orf}
G_{\mu\nu}&+\epsilon\Bigg[\varphi'(R_T)\left(\kappa T_{\mu\nu}-\frac{1}{2}\kappa T g_{\mu\nu}-\Lambda g_{\mu\nu}\right)\\
&-\frac{1}{2}\varphi(R_T) g_{\mu\nu}- [\nabla_\mu \nabla_\nu-g_{\mu\nu}\Box]\varphi'(R_T)\Bigg]=\kappa\, T_{\mu\nu}.\nonumber
\end{align}
where we have used eq.~(\ref{trace}) to help us express $R$ to the lowest order and
\be
R_T=-\kappa T-4\Lambda.
\ee

Let us derive the modified Friedmann equation corresponding to the order-reduced field eqs.~(\ref{orf}). Assuming a Friedmann-Lema\^itre-Robertson-Walker (FLRW) line element
\be
ds^2=-dt^2+a(t)^2 \left[\frac{dr^2}{1-k r^2}+r^2 d\theta^2+r^2\sin^2\theta d\phi^2\right],
\ee
where and $k=-1,0,1$ (hyperbolic, flat, hyperspherical) is the spatial curvature, and a perfect fluid description of matter for which
%\be
$T^{\mu\nu}=(\rho+p)u^\mu u^\nu+p g^{\mu\nu}$,
%\ee
where $\rho$ and $p$ are the energy density and pressure of the fluid respectively and $u^\mu$ the 4-velocity,
the zero-zero component of eq.~(\ref{orf}) yields
\begin{align}
\label{1f}
3H^2+\frac{3 k}{a^2}=&\left(1-\frac{1}{2}(1+3w) \epsilon \varphi'_T\right) \kappa\rho -\epsilon \varphi'_T \Lambda\nonumber\\
&-\frac{1}{2} \epsilon \varphi_T -3 \epsilon H\dot{\varphi'_T }
\end{align}
where $\varphi_T=\varphi(R_T)$, $\varphi'_T=\varphi'(R_T)$, $H\equiv \dot{a}/a$ and we have assumed a barotropic equation of state $p=w\rho$. Recalling now that energy is conserved in metric $f(R)$ gravity, we can use the equation $\dot{\rho}=-3H (1+w)\rho$ and the chain rule to express $\dot{\varphi'_T }$ in terms of $\rho$:
\be
\dot{\varphi'_T }=\varphi''(R_T) \frac{\partial R_T}{\partial T} \dot{T}=-3\kappa \varphi''_T(1+w)(1-3w)H\rho,
\ee
where $\varphi''_T=\varphi''(R_T)$. Replacing this back in eq.~(\ref{1f}) and using the $\epsilon=0$ value for $H^2$ in this term
one gets
\begin{align}
\label{orfried}
H^2=&\frac{1}{3}\kappa \rho -\frac{ k}{a^2} -\frac{\epsilon}{3}\Bigg[\frac{1}{2}(1+3w) \varphi'_T\kappa\rho + \varphi'_T \Lambda\\
&+\frac{1}{2}\varphi_T -3\left(\kappa\rho-\frac{3k}{a^2}\right)\kappa \varphi''_T(1+w)(1-3w)\rho\Bigg].\nonumber
\end{align}

Let as now find the condition for eq.~(\ref{orfried}) to lead to eq.~(\ref{lqf}). We first have to assume that the matter is actually a scalar field  ($w=1$) and that spacetime is spatially flat ($k=0$) in order to be in agreement with the approach leading to eq.~(\ref{lqf}). 
We also set $\Lambda=0$, as it is obvious that eq.~(\ref{lqf}) does not have a cosmological constant contribution. Eq.~(\ref{orfried}) becomes
\be
\label{orfried2}
H^2=\frac{1}{3}\kappa \rho-\frac{\epsilon}{3}\Bigg[2\varphi'_T\kappa\rho +\frac{1}{2}\varphi_T +12 \varphi''_T\kappa^2\rho^2\Bigg].
\ee
Suppose now that we would like to require eq.~(\ref{orfried2}) to be the same as some Friedmann equation with a modified source. The latter could be written without loss of generality as
\be
H^2=\frac{1}{3}\kappa\rho+\Psi(\rho)
\ee
where $\Psi$ is some algebraic function, and the requirement would be that $\varphi$ should satisfy
the differential equation
\be
\label{equal}
\epsilon\left(2\varphi'_T\kappa\rho 
+\frac{1}{2}\varphi_T +12\varphi''_T\kappa ^2\rho^2\right)=-3\Psi(\rho)
\ee
for a given $\Psi$. We are interested in eq.~(\ref{lqf}) here, so we take $\Psi=-\kappa \rho^2/(3\rho_c)$. Then, since $R_T=-2\kappa \rho$ and the prime denotes differentiation with respect to $R$, eq.~(\ref{equal}) is essentially the ordinary differential equation (ODE)
\be
\label{diffeq}
\frac{1}{2}\varphi(x)-x \frac{{\rm d}\varphi(x)}{{\rm d} x}+3 x^2 \frac{{\rm d}^2\varphi(x)}{{\rm d} x^2}=A x^2,
\ee
where $A=(4\epsilon \kappa \rho_c)^{-1}$. Clearly, $b x^2$ where $b$ is a constant is a specific solution of this equation, whereas it is easy to show that the homogeneous ODE has no solutions that are analytic functions of $x$, due to its specific structure (this is true for any $\Psi$ and, therefore, $\Psi$ is really what determines the solution)\footnote{For instance, one could use a power series to solve the homogeneous ODE: the fact that all independent terms would have exactly the same power would imply that all coefficients should be zero.}.  We, therefore, conclude that the general analytic solution is $\varphi(R)= (18 \epsilon \kappa \rho_c)^{-1}\,R^2$ or
\be
f(R)=R+\frac{\pi\gamma^3 l_p^2}{9\sqrt{3}}\,R^2+\ldots.
\ee
Clearly, the same procedure could have been followed for another function $\Psi$.  

We have, therefore, found a metric $f(R)$ action which, when treated as an effective action, leads to the desired Friedmann equations. It is important to note that the lagrangian $f$ we found is really an infinite series, as the dots indicate. Within the effective field theory framework we are allowed to remain agnostic on the rest of the terms and our calculation indicates that they should be subdominant to the $R^2$ correction in the curvatures under consideration.

On the other hand, it is well known that the gravitational lagrangian 
\be
\label{rsq}
L=R+a R^2 +b R^{\mu\nu} R_{\mu\nu}, 
\ee
leads to the same field equations with the lagrangian $L=R+c R^2$ for conformally flat metrics and for $c=a+b/3$
\cite{Barrow:1983rx}. Adding a term proportional to the Kretschmann scalar $R^{\mu\nu\alpha\beta}R_{\mu\nu\alpha\beta}$ on the other hand, is equivalent to a change in the coefficients $a$ and $b$, due to the the Gauss--Bonnet theorem. Therefore, the family of lagrangians of the form of eq~(\ref{rsq}) with $a+b/3=\pi\gamma^3 l_p^2/(9\sqrt{3})$ (plus other subdominant terms) will lead to the desired effective Friedmann equation. The same result could have been obtained if one followed the procedure we followed here for an a analytic function $f(R, R^{\mu\nu} R_{\mu\nu}, R^{\mu\nu\alpha\beta}R_{\mu\nu\alpha\beta})$ instead of simply $f(R)$.
One could also start immediately with the lagrangian (\ref{rsq}), show that it indeed leads to the desired result and then argue one dimensional grounds that it is the only (analytic) one. We chose the ``derivational" approach simply for demonstrative purposes. The fact that the effective action corresponding to eq.~(\ref{lqf}) is unique in the framework of metric $f(R)$ gravity, whereas there are is a whole family when it comes to more general actions, is merely because, due to the the high degree of symmetry assumed in their derivation,  Friedmann equations do not carry enough information to uniquely pinpoint a theory unless the latter is chosen from a restricted class.

To summarize, we have shown that a covariant effective action, or at least a family of covariant effective actions, which lead to the effective Friedmann equation as predicted from the massless scalar field model of LQC can be found in the framework of metric higher order theories of gravity. However, this requires the technique of order reduction, {\em i.e.}~it requires that the theory be treated as an effective field theory and that the extra degrees of freedom be considered spurious. The need for such an approach might be revealing some characteristics, for instance non-locality, which can be either intrinsic to the fundamental theory or  induced properties in the effective description due to some approximation.

Besides possibly revealing some generic characteristic of LQC, deriving a covariant action has certain other advantages. For instance it allows the study of the  phenomenology of the theory which goes beyond the effective Friedmann equation (but should be described by the effective action). In the specific action we find the corrections with respect to GR are Planck suppressed. Combining this with the fact that within our scheme the extra degrees of freedom are not to be considered, the low energy phenomenology should be indistinguishable from that of GR at the relevant length scales. However, there might still be differences, especially in the early universe \cite{Singh:2006sg}, which having full covariant field equation and an action might help reveal. A typical example would be cosmological perturbations.

Clearly, the effective action derived here corresponds to the simplified model of a scalar field and is, therefore,  valid in the range of validity of the approximation used to derive the effective hamiltonian for this model. However, the procedure described here can be used to derive effective actions for more general  effective Friedmann equations ({\em e.g.}~for a self interacting scalar \cite{Bojowald:2008ec}). Of course, it remains the objective of future investigations to determine whether the features of the effective covariant description found here will persist if other matter fields and more general spacetimes are considered  in LQC. 

A specific issue that deserves to be studied further is the following: In \cite{Chiou:2007sp} an anisotropic model was considered and effective Friedmann equations were derived that contained matter independent corrections which would persist in vacuum (see {\em e.g.}~appendix C of \cite{Chiou:2007sp}). The effective field theory technique used here cannot lead to any matter independent deviation from GR [see eq.~(\ref{orf})]. The same is true for the proposal of \cite{Olmo:2008nf} based on Palatini $f(R)$ gravity for a different reason: the latter reduces to GR plus a cosmological constant in vacuum. Since in anisotropic models one uses a deparameterization where the scalar field is taken as the time variable, the presence of matter is essential and there are technical differences between vacuum and non-vacuum models.  However, our previous observation excludes the presence of matter independent corrections even when matter is present. The absence of extra degrees of freedom from the  effective Friedmann equation seems to exclude the possibility of finding a remedy to this problem by considering theories of gravity with extra dynamical fields.. This seems to leave no available option for obtaining a covariant description for anisotropic models in LQC, unless one can obtain a different set of  effective equations by reconsidering the approach and the assumptions used in \cite{Chiou:2007sp} or is possibly willing to sacrifice background independence by  allowing for non-dynamical fields to be present in the effective action. This issue will be address in future work.

{\em Acknowledgments}: 
I am grateful to Ted Jacobson for a critical reading of this manuscript and 
several constructive suggestions and to Abhay Ashtekar and Martin Bojowald for enlightening discussions.
This work was supported by the NSF grant PHYS-0601800.

\end{document}